\documentclass{article}
\usepackage{geometry}
\usepackage{setspace}
\usepackage[dvipdfm]{graphicx}
\usepackage{graphics}
\usepackage{graphicx}
\usepackage{amsmath}
\usepackage{amsfonts}
\usepackage{amssymb}
\usepackage{multicol}
\ifx\pdfoutput\@undefined\usepackage[usenames,dvips]{color}
\else\usepackage[usenames,dvipsnames]{color}
\IfFileExists{pdfcolmk.sty}{\usepackage{pdfcolmk}}{} 
\fi
\usepackage[plainpages=false,pdfpagelabels,pagebackref=false,naturalnames=true,hyperindex=true,pdftitle={An algorithmic information-theoretic approach to the behaviour of financial markets},pdfauthor={Hector Zenil}]{hyperref}
\hypersetup{colorlinks=true, urlcolor=Cerulean,linkcolor=BrickRed,citecolor=RoyalBlue,a4paper, pdfpagemode=None, pdfstartview=FitH}
  
\begin{document}

\title{An algorithmic information-theoretic approach \\ to the behaviour of financial markets\footnote{Presented by Hector Zenil at the \emph{Workshop on Nonlinearity, Complexity and Randomness} under the title \emph{The Algorithmic Footprint in Empirical Data} in October of 2009 at the CIFREM/Department of Economics of the University of Trento, Italy.}}
\author{Hector Zenil$^{1,2}$ and Jean-Paul Delahaye$^2$ \\
$^{1}$IHPST, Universit\'e de Paris 1 (Panth\'eon-Sorbonne)\\
$^{2}$Laboratoire d'Informatique Fondamentale de Lille (USTL)\\
\href{mailto:hector.zenil@lifl.fr}{\{hector.zenil,delahaye\}@lifl.fr}}

\maketitle

\begin{abstract}
\noindent Using frequency distributions of daily closing price time series of several financial market indexes, we investigate whether the bias away from an equiprobable sequence distribution found in the data, predicted by algorithmic information theory, may account for some of the deviation of financial markets from log-normal, and if so for how much of said deviation and over what sequence lengths. We do so by comparing the distributions of binary sequences from actual time series of financial markets and series built up from purely algorithmic means. Our discussion is a starting point for a further investigation of the market as a rule-based system with an \emph{algorithmic} component, despite its apparent randomness, and the use of the theory of algorithmic probability with new tools that can be applied to the study of the market price phenomenon. The main discussion is cast in terms of assumptions common to areas of economics in agreement with an algorithmic view of the market. \\

\end{abstract}

\section{Introduction}

The main assumption regarding price modelling for option pricing is that stock prices in the market behave as stochastic processes, that is, that price movements are log-normally distributed. Unlike classical probability, algorithmic probability theory has the distinct advantage that it can be used to calculate the likelihood of certain events occurring based on their information content. We investigate whether the theory of algorithmic information may account for some of the deviation from log-normal of the data of price movements accumulating in a power-law distribution. 

We think that the power-law distribution may be an indicator of an information-content phenomenon underlying the market, and consequently that departures from log-normality can, given the accumulation of simple rule-based processes----a manifestation of hidden structural complexity----be accounted for by Levin's universal distribution, which is compatible with the distribution of the empirical data. If this is true, algorithmic probability could supply a powerful set of tools that can be applied to the study of market behaviour. Levin's distribution reinforces what has been empirically observed, viz. that some events are more likely than others, that events are not independent of each other, and that their distribution depends on their information content. Levin's distribution is not a typical probability distribution inasmuch as it has internal structure placing the elements according to their structure specifying their exact place in the distribution, unlike other typical probability distributions that may indicate where some elements accumulate without specifying the particular elements themselves.

The methodological discipline of considering markets as algorithmic is one facet of the algorithmic approach to economics laid out in \cite{vela}. The focus is not exclusively on the institution of the market, but also on agents (of every sort), and on the behavioural underpinnings of agents (rational or otherwise) and markets (competitive or not, etc.). 

We will show that the algorithmic view of the market as an alternative interpretation of the deviation from log-normal behaviour of  prices in financial markets is also compatible with some common assumptions in classical models of  market behaviour, with the added advantage that it points to the iteration of algorithmic processes as a possible cause of the discrepancies between the data and stochastic models. 

We think that the study of frequency distributions and the application of algorithmic probability could constitute a tool for estimating and eventually understanding the information assimilation process in the market, making it possible to characterise the information content of prices.

The paper is organised as follows: In \ref{traditional} a simplified overview of the basics of the stochastic approach to the behaviour of financial markets is introduced, followed by a section discussing the apparent randomness of the market. In section \ref{AITapproach}, the theoretic-algorithmic approach we are proposing herein is presented, preceded by a short introduction to the theory of algorithmic information, and followed by a description of the hypothesis testing methodology \ref{methodology}. In \ref{experiments}, tables of frequency distributions of price direction sequences for five different stock markets are compared to equiprobable (normal independent) sequences of length 3 and 4 to length 10 and to the output frequency distributions produced by algorithmic means. The alternative hypothesis, that is that the market has an algorithmic component and that algorithmic probability may account for the deviation of price movements from log-normality is tested, followed by a backtesting section \ref{backtesting} before introducing further considerations in \ref{furtherconsiderations} regarding common assumptions in economics. The paper ends with a short section that summarises conclusions and provides suggestions for further work in \ref{conclusions}.

\section{The traditional stochastic approach}
\label{traditional}

When events are independent of each other they accumulate in a normal (Gaussian) distribution. Stock price movements are for the most part considered to behave independently of each other. The random-walk like evolution of the stock market has motivated the use of Brownian motion for modelling price movements.

Brownian motion and financial modelling have been historically tied together\cite{cont}, ever since Bachelier proposed to model the price $S_t$ of an asset on the Paris stock market in terms of a random process of Brownian motion $W_t$ applied to the original price $S_0$. Thus $S_t = S_0 + \sigma W_t$. 

The process $S$ is sometimes called a \emph{log} (or \emph{geometric}) \emph{Brownian motion}. Data of price changes from the actual markets are actually too \emph{peaked} to be related to samples from normal populations. One can get a more convoluted model based on this process introducing or restricting the amount of randomness in the model so that it can be adjusted to some extent to account for some of the deviation of the empirical data to the supposedly log-normality.

A practical assumption in the study of financial markets is that the forces behind the market have a strong stochastic nature. The idea stems from the main assumption that market fluctuations can be described by classical probability theory. The multiplicative version of Bachelier's model led to the commonly used Black-Scholes model, where the log-price $S_t$ follows a random walk $S_t = S_0  exp[\sigma t + \sigma W_t]$.

\begin{figure}
  \begin{multicols}{2}
    \begin{center}
      \includegraphics[height=3.5cm]{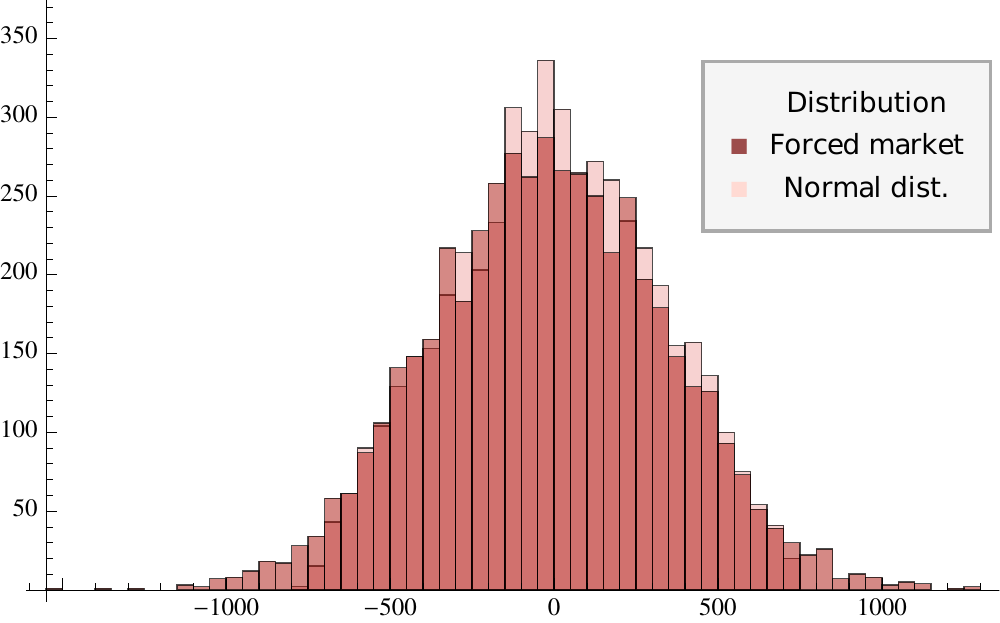}
  \caption{In a normal distribution any event is more or less like any other.}
       \includegraphics[height=3.5cm]{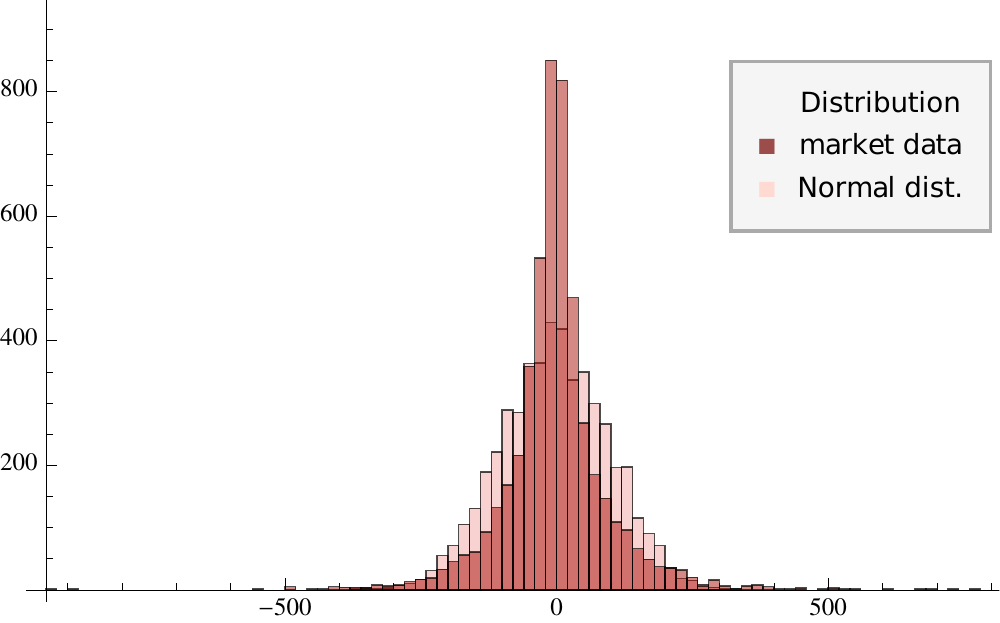}
  \caption{Events in a long-tailed (power-law) distribution indicate that certain days are not like any others. }
    \end{center}
  \end{multicols}
  \label{nonormal}
\end{figure}

\begin{figure}
  \begin{multicols}{2}
As found by Mandelbrot\cite{mandelbrot} in the 60s; prices do not follow a normal distribution; suggesting as it seems to be the case that some unexpected events happen more frequently than predicted by the Brownian motion model. On the right two of the plot were generated one by taking the central column of a rule 30 cellular automaton, and another by using the RandomInteger random number generator built in \emph{Mathematica}. Only one is the actual sequence of price movements for 3$\,$000 closing daily prices of the Dow Jones Index (DJI).

\begin{center}
\includegraphics[width=.45\textwidth]{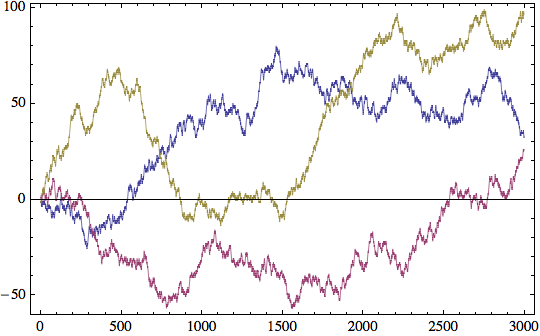}
\end{center}
\caption {Simulated Brownian motions together with a true segment of daily closing prices of the DJI.}
  \end{multicols}
\label{graphA}
\end{figure}

The kind of distribution in which price changes accumulate is a power-law in which high-frequency events are followed by low-frequency events, with the short and very quick transition between them characterised by asymptotic behaviour. Perturbations accumulate and are more frequent than if normally distributed, as happens in the actual market, where price movements accumulate in long-tailed distributions.  Such a distribution often points to specific kinds of mechanisms, and can often indicate a deep connection with other, seemingly unrelated systems.

\section{Apparent randomness in financial markets}
\label{wolfram}

The most obvious feature of essentially all financial markets is the apparent randomness with which prices tend to fluctuate. Nevertheless, the very idea of chance in financial markets clashes with our intuitive sense of the processes regulating the market. All processes involved seem deterministic. Traders do not only follow hunches but act in accordance with specific rules, and even when they do appear to act on intuition, their decisions are not random  but instead follow from the best of their knowledge of the internal and external state of the market. For example, traders copy other traders, or take the same decisions that have previously worked, sometimes reacting against information and sometimes acting in accordance with it. Furthermore, nowadays a greater percentage of the trading volume is handled electronically, by computing systems  (conveniently called algorithmic trading) rather than by humans. Computing systems are used for entering trading orders, for deciding on aspects of an order such as the timing, price and quantity, all of which cannot but be algorithmic by definition.

\emph{Algorithmic} however, does not necessarily mean \emph{predictable}. Several types of irreducibility, from non-computability to intractability to unpredictability, are entailed in most non-trivial questions about financial markets, as shown with clear examples in \cite{vela} and \cite{wolfram}.

In \cite{wolfram} Wolfram asks whether the market generates its own randomness, starting from deterministic and purely algorithmic rules. Wolfram points out that the fact that apparent randomness seems to emerge even in very short timescales suggests that the randomness (or a source of it) that one sees in the market is likely to be the consequence of internal dynamics rather than of external factors. In economists' jargon, prices are determined by endogenous effects peculiar to the inner workings of the markets themselves, rather than (solely) by the exogenous effects of outside events. 

Wolfram points out that pure speculation, where trading occurs without the possibility of any significant external input, often leads to situations in which prices tend to show more, rather than less, random-looking fluctuations. He also suggests that there is no better way to find the causes of this apparent randomness  than by performing an almost step-by-step simulation, with little chance of besting the time it takes for the phenomenon to unfold----the time scales of real world markets being simply too fast to beat. It is important to note that the intrinsic generation of complexity proves the stochastic notion to be a convenient assumption about the market, but not an inherent or essential one. 

\begin{figure}
  \begin{multicols}{2} Wolfram's proposal for modelling market prices would have a simple programme generate the randomness that occurs intrinsically. A plausible, if simple and idealised behaviour is shown in the aggregate to produce intrinsically random behaviour similar to that seen in price changes. In Figure 4, one can see that even in some of the simplest possible rule-based systems, structures emerge from a random-looking initial configuration with low information content. Trends and cycles are to be found amidst apparent randomness. 

\begin{center}
\includegraphics[width=.4\textwidth]{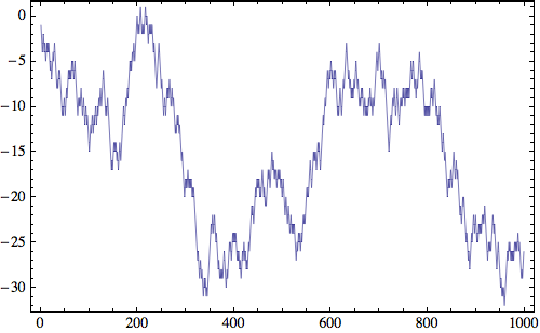}
\end{center}
\caption{Patterns out of nothing: random walk by 1$\,$000 data points generated using the \emph{Mathematica} pseudo-random number generator based on a deterministic cellular automaton.}
  \end{multicols}
\label{graphCb}
\end{figure}

\begin{figure}
  \begin{multicols}{2}
An example of a simple model of the market as shown in \cite{wolfram}, where each cell of a cellular automaton corresponds to an entity buying or selling at each step. The behaviour of a given cell is determined by the behaviour of its two neighbours on the step before according to a rule. The plot on the left gives as a rough analog of a market price differences of the total numbers of black and white cells at successive steps. A rule like rule 90 is additive, hence reversible, which means that it does not destroy any information and has 'memory' unlike the random walk model. Yet, due to its random looking behaviour, it is not trivial shortcut the computation or foresee any successive step. There is some \emph{randomness} in the initial condition of the cellular automaton rule that comes from outside the model, but the subsequent evolution of the system is fully deterministic. The way the series plot is calculated is written in \emph{Mathematica} as follows {\small Accumulate[Total/@(CA /. \{0$\rightarrow$-1\})]} {\normalsize with \emph{CA} the output evolution of rule 90 after 100 steps.}

\begin{center}
      \includegraphics[height=1.2cm]{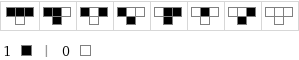}
      \includegraphics[height=5.3cm]{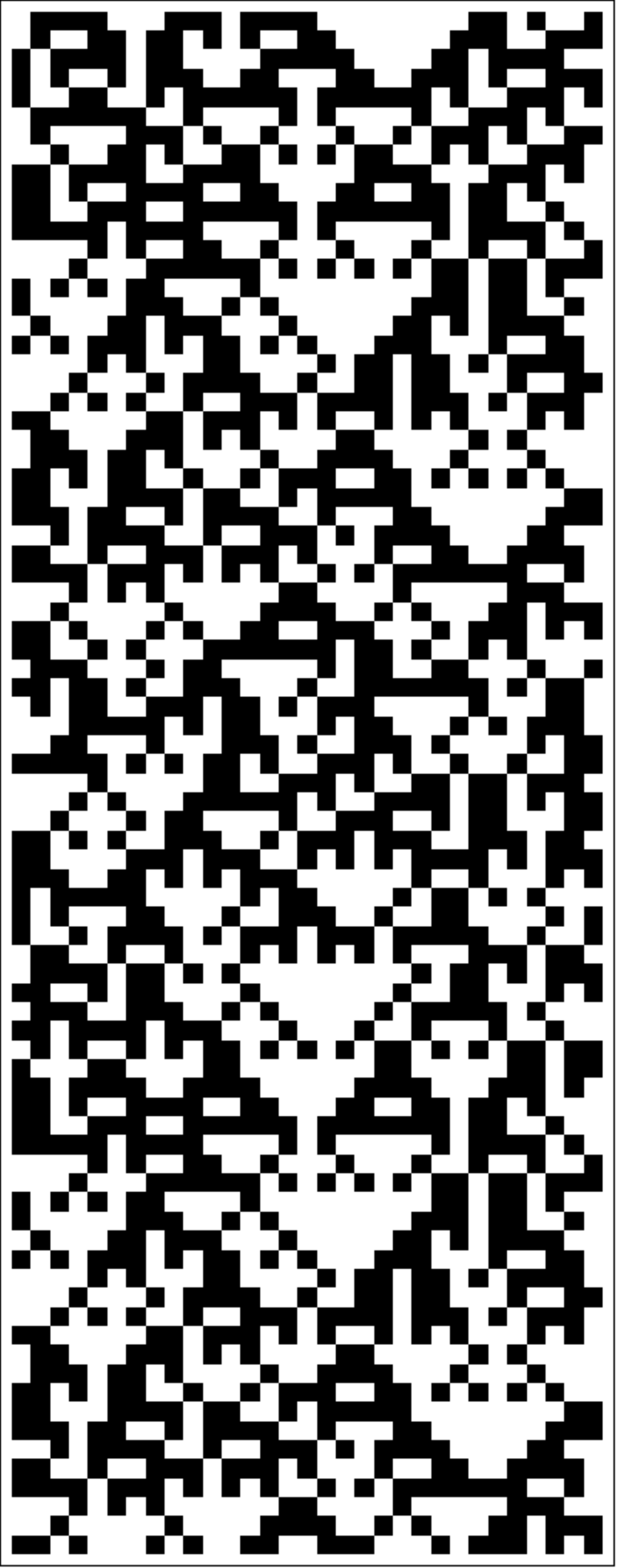}
      \includegraphics[height=2.4cm]{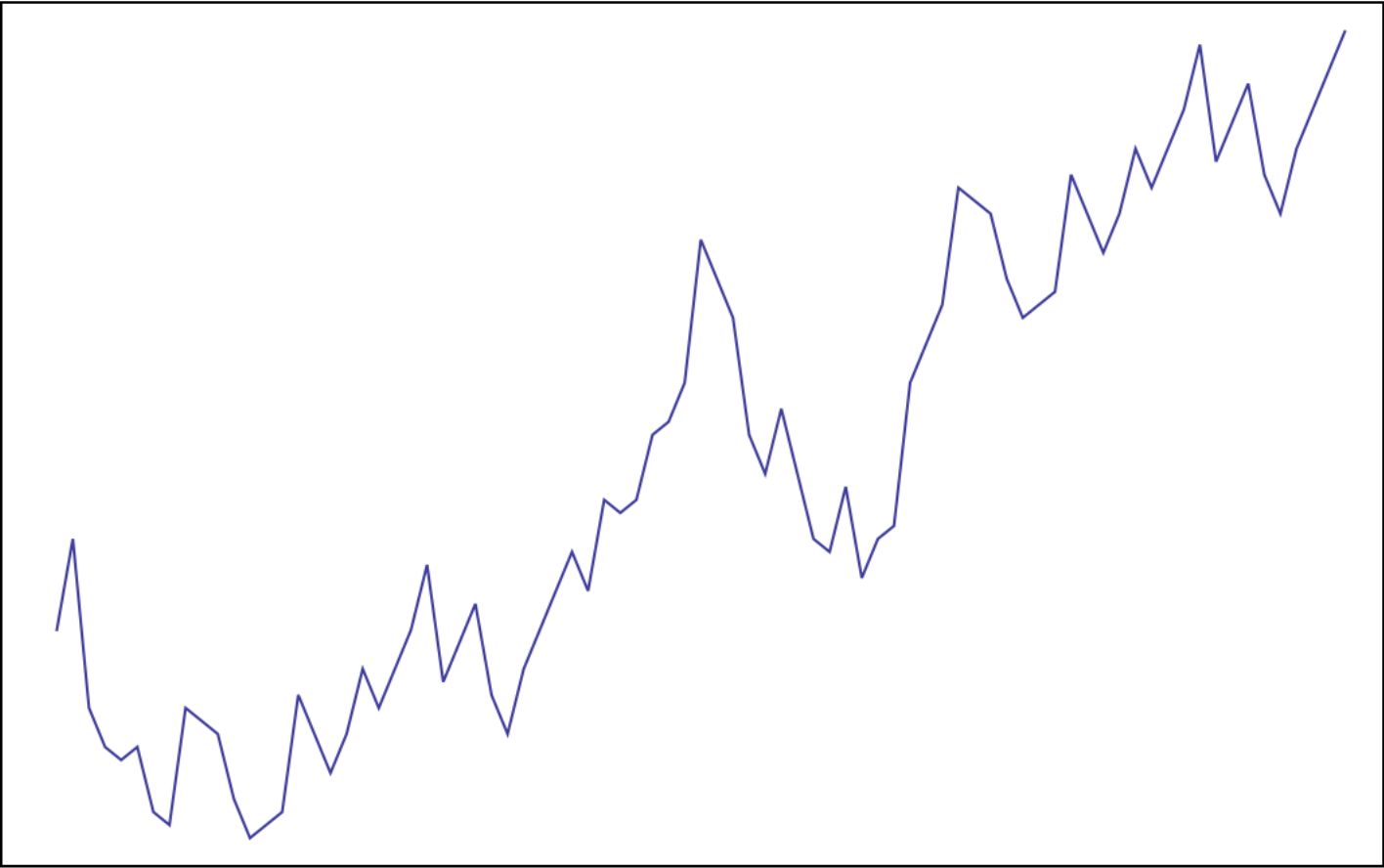}\\
  \caption{On the top, the rule 90 instruction table. On the left the evolution of rule 90 from a random init condition for 100 steps. On the right the total differences at every step between black and white cells.}
    \end{center}
  \end{multicols}
  \label{rule90}
\end{figure}

Economists may argue that the question is irrelevant for practical purposes. They are interested in decomposing time-series into a non-predictable and a presumably predictable signal in which they have an interest, what is traditionally called a trend. Whether one, both or none of the two signals is deterministic may be considered irrelevant as long as there is a part that is random-looking, hence most likely unpredictable and consequently worth leaving out.

What Wolfram's simplified model show, based on a simple rules, is that despite being so simple and completely deterministic, these models are capable of generating great complexity and exhibit (the lack of) patterns similar to the apparent randomness found in the price movements phenomenon in financial markets. Whether one can get the kind of crashes in which financial markets seem to cyclicly fall into depends on whether the generating rule is capable of producing them from time to time. Economists dispute whether crashes reflect the intrinsic instability of the market, or whether they are triggered by external events. In a model in \cite{lamper}, for example, sudden large changes are internally generated suggesting large changes are more predictable---both in magnitude and in direction as the result of various interactions between agents. If Wolfram's intrinsic randomness is what leads the market one may think one could then easily predict its behaviour if this were the case, but as suggested by Wolfram's Principle of Computational Equivalence it is reasonable to expect that the overall collective behaviour of the market would look complicated to us, as if it were random, hence quite difficult to predict despite being or having a large deterministic component.

Wolfram's Principle of Computational Irreducibility\cite{wolfram} says that the only way to determine the answer to a computationally irreducible question is to perform the computation. According to Wolfram, it follows from his Principle of Computational Equivalence (PCE) that ``\emph{almost all processes that are not obviously simple can be viewed as computations of equivalent sophistication: when a system reaches a threshold of computational sophistication often reached by non-trivial systems, the system will be computationally irreducible}.''

\section{An information-theoretic approach}
\label{AITapproach}

From the point of view of cryptanalysis, the algorithmic view based on frequency analysis presented herein may be taken as a hacker approach to the financial market. While the goal is clearly to find a sort of password unveiling the rules governing the price changes, what we claim is that the password may not be immune to a frequency analysis attack, because it is not the result of a true random process but rather the consequence of the application of a set of (mostly simple) rules. Yet that doesn't mean one can crack the market once and for all, since for our system to find the said password it would have to outperform the unfolding processes affecting the market---which, as Wolfram's PCE suggests, would require at least the same computational sophistication as the market itself, with at least one variable modelling the information being assimilated into prices by the market at any given moment. In other words, the market password is partially safe not because of the complexity of the password itself but because it reacts to the cracking method.

Whichever kind of financial instrument one looks at, the sequences of prices at successive times show some overall trends and varying amounts of apparent randomness. However, despite the fact that there is no contingent necessity of true randomness behind the market, it can certainly look that way to anyone ignoring the generative processes, anyone unable to see what other, non-random signals may be driving  market movements.

 Von Mises' approach to the definition of a random sequence, which seemed at the time of its formulation to be quite problematic, contained some of the basics of the modern approach adopted by Per Martin-L\"of\cite{lof}. It is during this time that the Keynesian\cite{keynes} kind of induction may have been resorted to as a starting point for Solomonoff's seminal work\cite{solomonoff} on algorithmic probability.

Martin-L\"of gave the first suitable definition of a random sequence. Intuitively, an algorithmically random sequence (or random sequence) is an infinite sequence of binary digits that appears random to any algorithm. This contrasts with the idea of randomness in probability. In that theory, no particular element of a sample space can be said to be random. Martin-L\"of randomness has since been shown to admit several equivalent characterisations in terms of compression, statistical tests, and gambling strategies.

The predictive aim of economics is actually profoundly related to the concept of predicting and betting. Imagine a random walk that goes up, down, left or right by one, with each step having the same probability. If the expected time at which the walk ends is finite, predicting that the expected stop position is equal to the initial position, it is called a martingale. This is because  the  chances of going up, down, left or right, are the same, so that  one ends up close to one's starting position,if not exactly at that position. In economics, this can be translated into a trader's experience. The conditional expected assets of a trader are equal to his present assets if a sequence of events is truly random.

Schnorr\cite{schnorr,downey} provided another equivalent definition in terms of martingales. The martingale characterisation of randomness says that no betting strategy implementable by any computer (even in the weak sense of constructive strategies, which are not necessarily computable) can make money betting on a random sequence. In a true random memoryless market, no betting strategy can improve the expected winnings, nor can any option cover the risks in the long term.

Over the last few  decades, several systems have shifted towards ever greater levels of complexity and information density. The result has been a shift towards Paretian outcomes, particularly within any event that contains a high percentage of informational content\footnote{For example, if one plots the frequency rank of words contained in a large corpus of text data versus the number of occurrences or actual frequencies, Zipf showed that one obtains a power-law distribution.}.

Departures from normality could be accounted for by the algorithmic component acting in the market,  as is consonant with some empirical observations and common assumptions in economics, such as rule-based markets and agents.

\begin{figure}
  \begin{multicols}{2}
If market price differences accumulated in a normal distribution, a rounding would produce sequences of 0 differences only. The \emph{mean} and the \emph{standard} \emph{deviation} of the market distribution are used to create a normal distribution, which is then subtracted from the market distribution. Rounding by the normal distribution cover, the elements in the tail are \emph{extracted} as shown in Figure 6.

    \begin{center}
      \includegraphics[height=3cm]{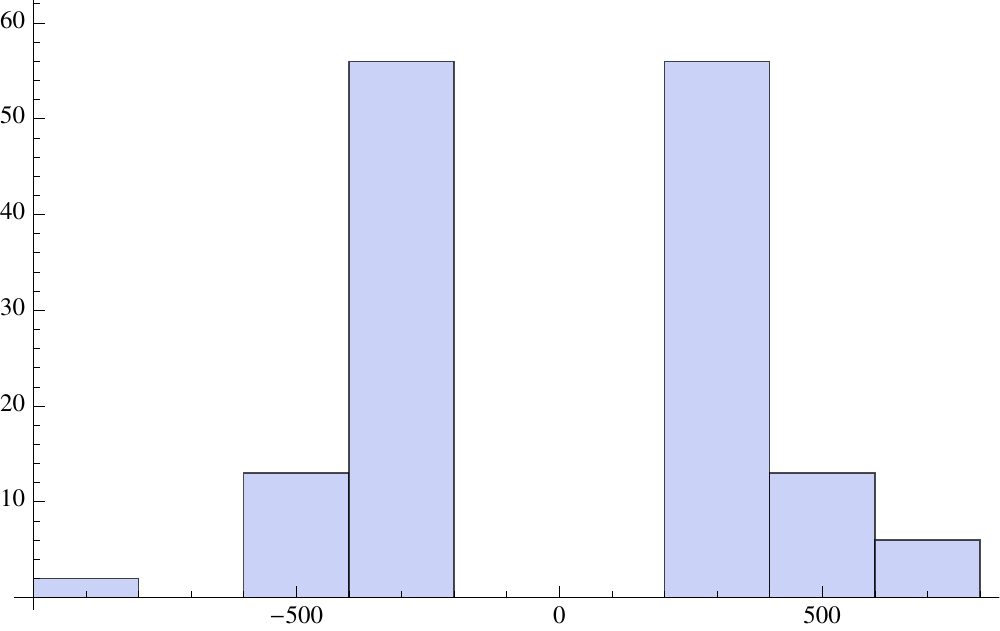}
  \caption{By extracting a normal distribution from the market distribution, the long-tail events are isolated.}
    \end{center}
  \end{multicols}
  \label{rule90}
\end{figure}

\subsection{Algorithmic complexity}

At the core of algorithmic information theory (AIT) is the concept of algorithmic complexity\footnote{Also known as programme-size complexity, or Kolmogorov complexity.}, a measure of the quantity of information contained in a string of digits. The algorithmic complexity of a string is defined as the length of the shortest algorithm that, when provided as input to a universal Turing machine or idealised simple computer, generates the string. A string has maximal algorithmic complexity  if the shortest algorithm able to generate it is not significantly shorter than the string itself, perhaps allowing  for a fixed additive constant. The difference in length between a string and the shortest algorithm able to generate it is the string's degree of compressibility. A string of low complexity is therefore highly compressible, as the information that it contains can be encoded in an algorithm much shorter than the string itself. By contrast, a string of maximal complexity is incompressible. Such a string constitutes its own shortest description: there is no more economical way of communicating the information that it contains than by transmitting the string in its entirety. In algorithmic information theory a string is algorithmically random if it is incompressible.

Algorithmic complexity is inversely related to the degree of regularity of a string. Any pattern in a string constitutes redundancy: it enables one portion of the string to be recovered from another, allowing a more concise description. Therefore highly regular strings have low algorithmic complexity, whereas strings that exhibit little or no pattern have high complexity.

The algorithmic complexity $K_U(s)$ of a string $s$ with respect to a universal Turing machine $U$ is defined as the binary length of the shortest programme $p$ that produces as output the string $s$.\\

\begin{center}
$K_U(s) = min\{|p|, U(p)=s)$
\end{center}

Algorithmic complexity conveys the intuition that a random string should be incompressible: no programme shorter than the size of the string produces the string.

Even though $K$ is uncomputable as a function, meaning that there is no effective procedure (algorithm) to calculate it, one can use the theory of algorithmic probability to obtain exact evaluations of $K(s)$ for strings $s$ short enough for which the halting problem can be solved for a finite number of cases due the size (and simplicity) of the Turing machines involved.

\subsection{Algorithmic probability}

What traders often end up doing in turbulent price periods is to leave aside the ''any day is like any other normal day'' rule, and fall back on their intuition, which leads to their unwittingly following a  model we believe to be better fitted to reality and hence to be preferred at all times, not just in times of turbulence. 

Intuition is based on weighting  past experience, with experience that is closer in time being more relevant. This is very close to the concept of algorithmic probability and the way it has been used (and was originally intended to be used\cite{solomonoff}) in some academic circles as a  theory of universal inductive inference\cite{hutter}.

Algorithmic probability assigns to objects an a priori probability that is in some sense universal. This a priori distribution has theoretical applications in a number of areas, including inductive inference theory and the time complexity analysis of algorithms. Its main drawback is that it is not computable and thus can only be approximated in practise.

The concept of algorithmic probability was first developed by Solomonof\cite{solomonoff} and formalised by Levin\cite{levin}. Consider an unknown process producing a binary string of length k bits. If the process is uniformly random, the probability of producing a particular string s is exactly $2^{-k}$, the same as for any other string of length $k$. Intuitively, however, one feels  that there should be a difference between a string that can be recognised and distinguished, and the vast majority of strings that are indistinguishable to us as  regards whether the underlying process is truly random.

Assume one tosses a fair coin $20$  three times and gets the following outcomes:

\begin{center}
00000000000000000000\\
01100101110101001011\\
11101001100100101101
\end{center}

the first outcome would be very unlikely because one would expect a patternless outcome from a fair coin toss, one that resembles the second and third outcomes. In fact, it would be far more likely  that a simple deterministic algorithmic process has generated this string. The same could be said for the market: one usually expects to see few if any patterns in its main indicators, mostly for the reasons set forth in section \ref{wolfram}. Algorithmic complexity captures this expectation of patternlessness by defining what a random-looking string looks like. On the other hand, algorithmic probability predicts that random-looking outputs are the exception rather than the rule when the generating process is algorithmic.

There is a measure which describes the expected output of an abstract machine when running a random programme. A process that produces a string $s$ with a programme $p$ when executed on a universal Turing machine U has probability $m(s)$. As $p$ is itself a binary string, $m(s)$ can be defined as being the probability that the output of a universal Turing machine\footnote{A universal Turing machine is an abstraction of a general-purpose computer. Essentially, as proven by Alan Turing, a universal computer can simulate any other computer on an arbitrary input by reading both the description of the computer to be simulated and the input thereof from its own tape.} $U$ is $s$ when provided with a sequence of fair coin flip inputs interpreted as a programme. \\

\begin{center}
\label{levindef}
$m(s) = \Sigma_{U(p) = s} 2^{-|p|} = 2^{-K(s) + O(1)}$
\end{center}

i.e. the sum over all the programmes $p$ for which the universal Turing machine $U$ outputs the string $s$ and halts.

Levin's universal distribution is so called because, despite being uncomputable, it has the remarkable property (proven by Leonid Levin himself) that among all the lower semi-computable semi-measures, it dominates every other\footnote{Since it is based on the Turing machine model, from which the adjective \emph{universal} derives, the claim depends on the Church-Turing thesis.}. This makes Levin's universal distribution the optimal prior distribution when no other information about the data is available, and the ultimate optimal predictor (Solomonoff's original motivation\cite{solomonoff} was actually to capture the notion of learning by inference) when assuming the process to be algorithmic (or more precisely, carried out by a universal Turing machine). Hence the adjective 'universal.'

The algorithmic probability of a string is uncomputable. One way to calculate the algorithmic probability of a string is to calculate the universal distribution by running a large set of abstract machines producing an output distribution, as we did in \cite{delahayezenil}.

\section{The study of the real time series vs. the simulation of an algorithmic market}

The aim of this work is to study of the direction and eventually the magnitude to time series of real financial markets. To that mean, we first develop a codification procedure translating financial series into binary digit sequences. Despite the convenience and simplicity of the procedure, the translation captures several important features of the actual behaviour of prices in financial markets. At the right level, a simplification of finite data into a binary language is always possible. Each observation measuring one or more parameters (e.g. price, trade name, etc.) is an enumeration of independent distinguishable values, a sequence of discrete values translatable into binary terms\footnote{Seeing it as a binary sequence may seem an oversimplification of the concept of a natural process and its outcome, but the performance of a physical experiment always yields data written as a sequence of individual observations sampling certain phenomena.}.

\subsection{From AIT back to the behaviour of financial markets}

Different market theorists will have different ideas about the likely pattern of 0s and 1s that can be expected from a sequence of price movements. Random walk believers would favour random-looking sequences in principle. Other analysts may be more inclined to believe that patterned-looking sequences can be spotted in the market, and may attempt to describe and exploit these patterns, eventually deleting them. 

In an early anticipation of an application of AIT to the financial market \cite{mansilla}, it was reported that the information content of price movements and magnitudes seem to drastically vary when measured right before crashes compared to periods where no financial turbulence is observed. As described in \cite{mansilla}, this means that sequences corresponding to critical periods show a qualitative difference compared to the sequences corresponding to periods where stable and modelled by the traditional stochastic models and when the information content of the market is very low (looks random as carrying no information). In \cite{mansilla}, the concept of conditional algorithmic complexity is used to measure this differences in the time series of price movements in two different financial markets (NASDAQ and the Mexican IPC).

We will analyse the complexity of a sequence $s$ of encoded price movements, as described in the section \ref{experiments}. We will see whether this distribution approaches one produced artificially---by means of algorithmic processes---in order to conjecture the algorithmic forces at play in the market, rather than simply assume a pervasive randomness. Exploitable or not, we think that  price movements may have an algorithmic component, even if some of this complexity is disguised behind apparent randomness.

According to Levin's distribution, in a world of computable processes, patterns which result from simple processes are relatively likely, while patterns that can only be produced by very complex processes are relatively unlikely. Algorithmic probability would predict, for example, that consecutive runs of the same magnitude, i.e. runs of pronounced falls and rises, and runs of alternative regular magnitudes have greater probability than random-looking changes. If one fails to  discern the same simplicity in the market as is to be observed in certain  other real world data sources\cite{zenildelahaye}, it is likely due to the dynamic of the stock market, where the exploitation of any regularity to make a profit results in the deletion of that  regularity. Yet these regularities may drive the market and may be detected upon closer examination. For example, according to the classical theory, based on the average movement on a random walk, the probability of strong crashes is nil or very low. Yet in actuality they occur in cycles over and over.

What is different in economics is the nature of the dynamics some of the data is subject to, as discussed in section \ref{wolfram}, which underscores the fact that patterns are quickly erased by economic activity itself, in the search for an economic equilibrium (e.g. the stock market).

Assuming an algorithmic hypothesis, that is that there is a rule-based---as opposed to a purely stochastic---component in the market, one could apply the tools of the theory of algorithmic information, just as assuming random distributions led to the application of the traditional machinery of probability theory.

If this algorithmic hypothesis were true, the theory says that Levin's distribution is the optimal predictor. In other words, one could run a large number of machines to simulate the market, and $m$, the algorithmic probability based on Levin's universal distribution would provide some insights into the particular direction and magnitude of a price based on the fact that the market has a rule-based element. The correlation found in the experiments described in the next section \ref{experiments} suggests that Levin's distribution may turn out to be a way to calculate and approximate this potentially algorithmic segment in the market.

\subsection{Unveiling the machinery}

When observing a certain phenomenon, its outcome $f$ can be seen as the result of a process $P$. One can then ask what the probability distribution of $P$ generating $f$ looks like. A probability distribution of a process is a description of the relative number of times each possible outcome occurs in a number of trials.

In a world of computable processes, Levin's semi-measure (a.k.a universal distribution) establishes that patterns which result from simple processes (short programmes) are likely, while patterns produced by complicated processes (long programmes) are relatively unlikely. Unlike other probability measures, Levin's semi-measure (denoted by $m$) is not only a probability distribution establishing that there are some objects that have a certain probability of occurring according to said distribution, it is also a distribution specifying the order of the particular elements in terms of their individual information content.

Figure 7 suggests that by looking at the behaviour of one market, the behaviour of the others may be predicted.  But this cannot normally be managed quickly enough for the information to be of any actual use  (in fact the very intention of succeeding in one market by using information from another may be the cause rather than the consequence of the correlation).

\begin{figure}
  \begin{multicols}{2}

It is not only in  times of great volatility that one can see that markets are correlated to each other. This correlation means that, as may be expected, markets systematically react to each other. One can determine the information assimilation process time by looking at the correlations of sequences of daily closing prices of different lengths for five of the largest European and U.S. stock markets. It is evident that they react neither immediately nor after an interval of several days.  As suggested by the table \ref{correlationm}, over a period of 20 years, from January 1990 to January 2010, the average assimilation time is about a week to a week and a half. For one thing, the level of confidence of the correlation confirms that even if some events may be seen as randomly produced, the reactions of the markets follow each other and  hence are neither  independent of each other nor completely random. The correlation matrix \ref{correlationm} exhibits the Spearman rank correlation coefficients, followed by the number of elements compared (number of sequence lengths found in one or another  market ), underlining the significance of the correlation between them.

    \begin{center}
      \includegraphics[height=4cm]{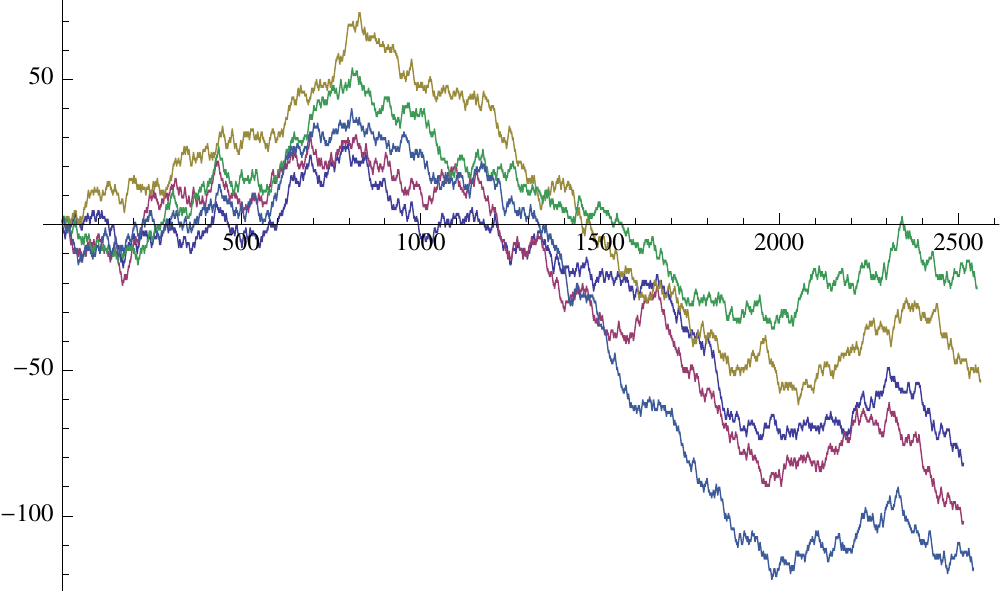}
  \caption{Series of daily closing prices of five of the largest stock markets from 01/01/2000 to January 01/01/2010. The best sequence length correlation suggests that markets catch up with each other (assimilate each others'  information) in about 7 to 10 days on average.}
    \end{center}
  \end{multicols}
  \label{walks}
\end{figure}

In the context of economics, if we accept the algorithmic hypothesis (that price changes are algorithmic, not random), $m$ would provide the algorithmic probability of a certain price change  happening, given the history of the price. An anticipation of the use of this prior distribution as an inductive theory in economics is to be found in \cite{vela}. But following that model would require us to calculate the prior distribution $m$, which we know is uncomputable. We proceed by approaching $m$ experimentally in order to show what the distribution of an algorithmic market would look like, and eventually use it in an inductive framework.

Once $m$ is approximated, it can be compared to the distribution of the outcome in the real world (i.e. the empirical data on  stock market price movements). If the output of a process approaches a certain probability distribution, one accepts, within a reasonable degree of statistical certainty, that the generating process is of the nature suggested by the distribution. If it is observed that an outcome $s$ occurs with probability $m(s)$, and the data distribution approaches $m$, one would be persuaded, within the same degree of certainty, to accept a uniform distribution as the footprint of a random process (where events are independent of each other), i.e. that the source of the data is suggested by the distribution $m$.

What Levin's distribution implies is that most rules are simple because they produce simple patterned strings, which algorithmically complex  rules are unlikely to do. A simple rule, in terms of algorithmic probability, is the average kind of rule producing a highly frequent string, which according to algorithmic probability has a low random complexity (or high organised complexity), and therefore looks patterned. This is the opposite of what a complex rule would be, when defined in the same terms---it  produces a pattern-less output, and is hence random-looking. 

The outcomes of simple rules have short descriptions because they are less algorithmically complex, means that in some sense \emph{simple} and \emph{short} are connected, yet large rules may also be simple despite not being short in relative terms. And the outcomes of the application of simple rules tend to accumulate exponentially faster than the outcomes of complicated rules. This causes some events to occur more often than others and therefore to be dependant on each other. Those events happening more frequently will tend to drastically outperform  other events and will do so by quickly beginning to follow the irregular pattern and doing so closely for a while. 

The first task  is therefore to produce a distribution by purely algorithmic means using abstract computing machines\footnote{One would actually need to think of one-way non-erasing Turing machines to produce a suitable distribution analogous to what could be expected from a sequence of events that have an unrewritable past and a unique direction (the future), but the final result shouldn't be that different according to the theory.}---by running abstract computational devices like Turing machines and cellular automata.

\subsection{Binary encoding of the market direction of prices}

Market variables have three main indicators. The first is whether there is a significant price change (i.e. larger than, say, the random walk expectation), the second is the direction of a price change (rising or falling), and thirdly, there is its magnitude. We will focus our first attempt on the direction of price changes, since this may be the most valuable of the three (after all whether one is likely to make or lose money is the first concern, before one ponders the magnitude of the possible gain or loss).

In order to verify that the market carries the algorithmic signal, the information content of the non-binary tuples can be collapsed to binary tuples. One can encode the price change in a single bit by 'normalizing' the string values, with the values of the entries themselves losing direction and magnitude but capturing price changes.

Prices are subject to such strong forces (interests) in the market that it would be naive to think that they could remain the same to any decimal fraction of precision, even if they were meant to remain the same for a period of time. In order to spot the algorithmic behaviour one has to provide some stability to the data by getting rid of precisely the kind of minor fluctuations that a random walk may predict. If not,  one notices strong biases disguising the real patterns. For example, periods of price fluctuations would appear less likely than  they are in reality if one allows decimal fluctuations to count as much as any other fluctuation\footnote{The same practise is common in time-series decomposition analysis, where the sole focus of interest is the average movement, in order that trends, cycles or other potential 
regularities may be discerned.}.

This is because from one day to another the odds that prices will remain exactly the same up to the highest precision is extremely unlikely, due to the extreme forces and time exposure  they are subject to. Even though it may seem that at such a level of precision one or more decimal digits could represent a real price change, for most practical purposes there is no qualitative change when prices close to the hundreds, if not actually in the hundreds, are involved. Rounding the price changes to a decimal fraction provides some stability, e.g. by improving the place of the $0^n$ tuple towards the top, because as we will see, it turns out to be quite well placed at the top once the decimal fluctuations have been gotten rid of. One can actually think of this as using the Brownian motion expectation to get rid of meaningless changes. In other words, the Brownian movement is stationary in our analysis, being the predictable (easy) component of the minor fluctuations, while the actual objects of study are the wilder dynamics of larger fluctuations. We have changed the focus from what analysts are used to considering as the signal, to noise, and recast what they consider noise as the  algorithmic footprint in empirical data.

\subsubsection{Detecting rising vs. falling}
\label{detecting}

As might be expected given the apparent randomness of the market and the dynamics to which it is subject, it would be quite difficult to discern clear patterns of rises vs. falls in market prices.

The asymmetry in the rising and falling ratio is explained by the pattern deletion dynamic. While slight rises and falls have the same likelihood of occurring, making us think they follow a kind of random walk bounded by the average movement of Brownian motion, significant rises are quickly deleted by people taking advantage of them, while strong drops quickly worsen because of people trying to sell rather than buy at a bargain. This being so, one expects to see longer sequences of drops  than of rises, but actually one sees the contrary, which suggests that periods of optimism are actually longer than periods of pessimism, though periods of pessimism are stronger in terms of price variation. In \cite{mansilla} it was reported that the information content of price movements and magnitudes seem to drastically vary when measured to intervals of high volatility (particularly right before crashes) compared to periods where no financial turbulence is observed.

We construct a binary series associated to each real time series of financial markets as follows. Let $\{p_t\}$ be the original time series of daily closing prices of a financial market for a period of time $t$. Then the binary $n-1$ elements of the time series $\{b_t\}$ are the price differences calculated as follows:

\begin{displaymath}
   b_n = \left\{
     \begin{array}{lr}
       1 & p_{\text{n+1}} > p_n \\
       0 & p_{\text{n+1}} \leq p_n
     \end{array}
   \right.
\end{displaymath} 
\\
The frequency of binary tuples of length $n$ from $\{b_t\}$ will be compared to the frequency of binary tuples of length $n$ obtained by running abstract machines (deterministic Turing machines and one-dimensional cellular automata).

\subsection{Calculating the algorithmic time series}
\label{methodology}

Constructing Levin's distribution $m$ from abstract machines is therefore necessary in order to strengthen the algorithmic hypothesis. In order to make a meaningful comparison with what can be observed in a purely rule-governed market, we will construct from the ground up an experimental distribution by running algorithmic machines (Turing machines and cellular automata). An abstract machine consists of a definition in terms of input, output, and the set of allowable operations used to turn the input into the output. They are of course algorithmic by nature (or by definition).

The Turing machine model represents the basic framework underlying many concepts in computer science, including the definition of algorithmic complexity. In this paper we use the output frequency distribution produced by running a finite, yet very large set of Turing machines with empty input as a means to approximate $m$. There are 11$\,$019$\,$960$\,$576 four-state two-symbol Turing machines (for which the halting condition is known, thanks to the Busy Beaver game). The conception of the experiment and further details are provided in \cite{delahayezenil} and \cite{zenildelahaye}. These Turing machines produced an output, from which a frequency distribution of tuples was calculated and compared to the actual time series from the market data produced by the stock markets encoded as described in \ref{detecting}. A forthcoming paper provides more technical details\cite{delahayezenil2}.

Also, a frequency distribution from a sample of 10$\,$000 four-color totalistic cellular automata was built from a total of 1$\,$048$\,$575 possible four-color totalistic cellular automata, each rule starting from an random initial configuration of 10 to 20 black and white cells, and running for 100 steps  (hence arbitrarily halted). Four-color totalistic cellular automata produce four-symbol sequences. However only the binary were taken into consideration, with the purpose of building a binary frequency distribution. The choice of this cellular automata (CA) space was dictated by the fact that the smallest CA space is too small, and the next smallest space too large to extract a significant enough sample from it. Having chosen a sample of four-color totalistic CA, the particular rules sample was randomly generated.

For all the experiments, stock market datasets (of daily closing prices)  used covered the same period of time: from January 1990 through January 2010. The stock market code names can be readily connected with their full index names or symbols.

Spearman's rank coefficient was the statistical measure of the correlation between rankings of elements in the frequency distributions. As is known, if there are no repeated data values, a perfect Spearman correlation of $+1$ or $-1$ occurs when each of the variables is a perfect monotone function of the other. While 1 indicates perfect correlation (i.e. in the exact order), $-1$ indicates perfect negative correlation (i.e. perfect inverse order).

\newpage

\section{Experiments and Results}
\label{experiments} 

It is known that for any finite series of a sequence of integer values, there is a family of countable infinite computable functions that fit the sequence over its length. Most of them will be terribly complicated, and any attempt to find the simplest accounting for the sequence will face uncomputability. Yet using the tools of algorithmic information theory, one can find the simplest function for short sequences by deterministic means, testing a set of increasingly  complex functions starting from the simplest and proceeding until the desired sequence assuring the simplest fit is arrived at. Even if the likelihood of remaining close to the continuation of the series remains low, one can recompute the sequence and find good approximations for short periods of time. 

The following experiment uses the concept of algorithmic complexity to find the best fit in terms of the simplest model fitting the data, assuming the source to be algorithmic (rule-based).

\subsection{Rankings of price variations}

In the attempt to capture the price change phenomenon in the stock market, we have encoded price changes in binary strings. The following are tables capturing the rise and fall of prices and their occurrence in a ranking classification.

\subsubsection{Carrying an algorithmic signal}

Some regularities and symmetries in the sequence distribution of price directions in the market may be accounted for by an algorithmic signal, but they also follow from a normal distribution. For example, the symmetry between the left and right sides of the Gaussian curve with zero skewness means that reverted and inverted strings (i.e. consecutive runs of events of prices going up or down) will show similar frequency values, which is also in keeping with the intuitive expectation of the preservation of complexity invariant to certain basic transformations (a sequence of events 1111$\ldots$ are equally likely to occur as 0000$\ldots$, or 0101$\ldots$ and 1010$\ldots$).

This means that one would expect $n$ consecutive runs of rising prices to be as likely as n consecutive runs of falling prices. Some symmetries, however, are broken in particular scenarios. In the stock market for example, it is very well known that sequences of drastic falls are common from time to time, but never sequences of drastic price increases, certainly not increases of the same magnitude as the worst price drops. And we witnessed such a phenomenon in the distributions from the DJI. Some other symmetries may be accounted for by business cycles engaged in the search for economic equilibrium.

\subsubsection{Correlation matrix: market vs. market}
\label{correlationm}

$
\begin{array}{|c|c|c|c|c|c|c|c|}
\hline
\text{market vs. market} & 4 & 5 & 6 & 7 & 8 & 9 & 10 \\
\hline
\text{CAC40 vs. DAX} & 0.059|16 & 0.18|32 & 0.070|62 & 0.37|109 & 0.48|119 & 0.62|87 & 0.73|55 \\
 \text{CAC40 vs. DJIA} & 0.31|16 & 0.25|32 & 0.014|62 & 0.59|109 & 0.27|124 & 0.34|95 & 0.82|51 \\
 \text{CAC40 vs. FTSE350} & 0.16|16 & -0.019|32 & -0.18|63 & 0.15|108 & 0.59|114 & 0.72|94 & 0.73|62 \\
 \text{CAC40 vs. NASDAQ} & 0.30|16 & 0.43|32 & 0.056|63 & 0.16|111 & 0.36|119 & 0.32|88 & 0.69|49 \\
 \text{CAC40 vs. SP500} & 0.14|16 & 0.45|31 & -0.085|56 & -0.18|91 & 0.16|96 & 0.49|73 & 0.84|45 \\
 \text{DAX vs. DJIA} & 0.10|16 & -0.14|32 & 0.13|62 & 0.37|110 & 0.56|129 & 0.84|86 & 0.82|58 \\
 \text{DAX vs. FTSE350} & 0.12|16 & -0.029|32 & 0.12|63 & 0.0016|106 & 0.54|118 & 0.81|89 & 0.80|56 \\
 \text{DAX vs. NASDAQ} & 0.36|16 & 0.35|32 & 0.080|62 & 0.014|110 & 0.64|126 & 0.55|96 & 0.98|48 \\
 \text{DAX vs. SP500} & 0.38|16 & 0.062|31 & -0.20|56 & -0.11|88 & 0.11|94 & 0.43|76 & 0.63|49 \\
 \text{DJIA vs. FTSE350} & 0.35|16 & -0.13|32 & -0.022|63 & 0.29|107 & 0.57|129 & 0.76|99 & 0.86|56 \\
 \text{DJIA vs. NASDAQ} & -0.17|16 & -0.13|32 & 0.0077|62 & 0.079|112 & 0.70|129 & 0.57|111 & 0.69|64 \\
 \text{DJIA vs. SP500} & -0.038|16 & 0.32|31 & -0.052|55 & 0.14|89 & 0.37|103 & 0.32|86 & 0.60|59 \\
 \text{FTSE350 vs. NASDAQ} & 0.36|16 & 0.38|32 & -0.041|63 & 0.54|108 & 0.68|126 & 0.57|107 & 0.66|51 \\
 \text{FTSE350 vs. SP500} & 0.50|16 & 0.50|31 & 0.12|56 & -0.11|92 & 0.25|101 & 0.26|96 & 0.29|66 \\
 \text{NASDAQ vs. SP500} & 0.70|16 & 0.42|31 & 0.20|56 & 0.024|91 & 0.41|111 & 0.23|102 & 0.42|61\\
 \hline
\end{array}
$
\begin{center} Table 1. Stock market vs. stock market correlations.
\end{center}

With algorithmic probability in hand, one may predict that alternations and consecutive events of the same type and magnitude are more likely, because they may be algorithmically more simple. One may, for example, expect to see symmetrical events occurring more often, with reversion and complementation occurring together in groups (i.e. a string $10^n$ occurring together with $0^n1$ and the like). In the long-term, business cycles and economic equilibria may  also be explained in information theoretic terms, because for each run of events there are the two complexity-preserving symmetries, reversion and complementation, that always follow their counterpart sequences (the unreversed and complement of the complement), producing a cyclic type of behaviour.

The correlations shown in table \ref{correlationm2} indicate what is already assumed in looking for cycles and trends, viz. that these underlying cycles and trends in the markets are more prominent when \emph{deleting} Brownian \emph{noise}. As shown below, this may be an indication that the tail of the distribution has a stronger correlation than the elements covered by the normal curve, as could be inferred from the definition of a random walk (that is, that random walks are not correlated at all). 

The entries in each comparison table consist of the Spearman coefficient followed by the number of elements compared. Both determine the level of confidence and are therefore essential for estimating the correlation. Rows compare different stock markets over different sequence lengths of daily closing prices, represented by the columns. It is also worth noting when the comparison tables, such as \ref{correlationm2}, have no negative correlations.

\subsubsection{Trend correlation matrix: rounded market vs. rounded market}
\label{correlationm2}

$
\begin{array}{|c|c|c|c|c|c|c|c|}
\hline
\text{market vs. market} & 4 & 5 & 6 & 7 & 8 & 9 & 10 \\
\hline
 \text{CAC40 vs. DAX} & 0.58|11 & 0.58|15 & 0.55|19 & 0.37|26 & 0.59|29 & 0.55|31 & 0.60|28 \\
 \text{CAC40 vs. DJIA} & 0.82|11 & 0.28|15 & 0.28|21 & 0.29|24 & 0.52|29 & 0.51|32 & 0.33|36 \\
 \text{CAC40 vs. FTSE350} & 0.89|11 & 0.089|15 & 0.41|20 & 0.17|22 & 0.59|30 & 0.28|28 & 0.30|34 \\
 \text{CAC40 vs. NASDAQ} & 0.69|11 & 0.27|14 & 0.28|18 & 0.44|23 & 0.30|30 & 0.17|34 & 0.61|30 \\
 \text{CAC40 vs. SP500} & 0.85|11 & 0.32|15 & 0.49|20 & 0.55|24 & 0.42|33 & 0.35|35 & 0.34|36 \\
 \text{DAX vs. DJIA} & 0.76|11 & 0.45|16 & 0.56|20 & 0.35|26 & 0.34|28 & 0.25|35 & 0.24|33 \\
 \text{DAX vs. FTSE350} & 0.61|11 & 0.30|16 & 0.58|19 & 0.14|25 & 0.30|29 & 0.34|30 & 0.21|31 \\
 \text{DAX vs. NASDAQ} & 0.40|11 & 0.27|16 & 0.36|18 & 0.75|25 & 0.28|29 & 0.28|35 & 0.50|33 \\
 \text{DAX vs. SP500} & 0.14|12 & 0.36|17 & 0.72|20 & 0.64|28 & 0.42|31 & 0.52|34 & 0.51|32 \\
 \text{DJIA vs. FTSE350} & 0.71|11 & 0.30|16 & 0.63|20 & 0.71|22 & 0.21|28 & 0.28|31 & 0.35|33 \\
 \text{DJIA vs. NASDAQ} & 0.58|11 & 0.52|15 & 0.33|19 & 0.58|23 & 0.46|29 & 0.49|37 & 0.51|35 \\
 \text{DJIA vs. SP500} & 0.70|11 & 0.20|16 & 0.45|21 & 0.29|26 & 0.35|32 & 0.37|36 & 0.55|36 \\
 \text{FTSE350 vs. NASDAQ} & 0.73|11 & 0.57|15 & 0.70|17 & 0.48|23 & 0.62|28 & 0.34|33 & 0.075|35 \\
 \text{FTSE350 vs. SP500} & 0.66|11 & 0.65|16 & 0.56|19 & 0.18|24 & 0.64|32 & 0.32|32 & 0.52|38 \\
 \text{NASDAQ vs. SP500} & 0.57|11 & 0.37|15 & 0.41|18 & 0.32|24 & 0.30|34 & 0.19|35 & 0.35|40\\
 \hline
\end{array}
$
\begin{center} Table 2. In contrast, when the markets are compared to random price movements, which accumulate in a normal curve, they exhibit no correlation or only a very weak correlation, as shown in \ref{correlationm3}:
\end{center}

\subsubsection{Correlation matrix: market vs. random}
\label{correlationm3}

$
\begin{array}{|c|c|c|c|c|c|c|c|}
\hline
 \text{r. market vs. random} & 4 & 5 & 6 & 7 & 8 & 9 & 10 \\
\hline
 \text{DJIA vs. random} & -0.050|16 & 0.080|31 & -0.078|61 & 0.065|96 & 0.34|130 & 0.18|120 & 0.53|85 \\
 \text{SP500 vs. random} & 0.21|16 & -0.066|30 & 0.045|54 & -0.16|81 & 0.10|99 & 0.29|87 & 0.32|57 \\
 \text{NASDAQ vs. random} & 0.12|16 & -0.095|31 & 0.11|60 & 0.14|99 & 0.041|122 & 0.29|106 & 0.57|68 \\
 \text{FTSE350 vs. random} & 0.16|16 & -0.052|31 & 0.15|61 & 0.14|95 & 0.30|122 & 0.50|111 & 0.37|77 \\
 \text{CAC40 vs. random} & 0.32|16 & -0.15|31 & -0.13|60 & 0.16|99 & 0.19|119 & 0.45|109 & 0.36|78 \\
 \text{DAX vs. random} & 0.33|16 & 0.023|31 & 0.20|60 & 0.14|95 & 0.26|129 & 0.31|104 & 0.31|77 \\
\hline
\end{array}
$
\begin{center} Table 3. When random price movements are compared to rounded prices of the market (denoted by ''r. market'' to avoid accumulation in the normal curve) the correlation coefficient is too weak, possessing no significance at all. This may indicate that it is the prices behaving and accumulating in the normal curve that effectively lead the overall correlation.
\end{center}

\subsubsection{Correlation matrix: rounded market vs. random}
\label{correlationm4}

$
\begin{array}{|c|c|c|c|c|c|c|c|}
 \hline
  \text{market vs. random} & 4 & 5 & 6 & 7 & 8 & 9 & 10 \\
 \hline
 \text{DJIA vs. random} & 0.21|12 & -0.15|17 & 0.19|23 & -0.033|28 & -0.066|33 & 0.31|29 & 0.64|15 \\
 \text{SP500 vs. random} & -0.47|12 & -0.098|17 & -0.20|25 & 0.32|31 & 0.20|38 & 0.41|29 & 0.32|20 \\
 \text{NASDAQ vs. random} & -0.55|11 & -0.13|16 & -0.093|20 & 0.18|26 & 0.015|37 & 0.30|35 & 0.38|25 \\
 \text{FTSE350 vs. random} & -0.25|11 & -0.24|16 & -0.053|22 & -0.050|24 & -0.11|31 & 0.25|23 & 0.49|13 \\
 \text{CAC40 vs. random} & -0.12|11 & -0.14|15 & 0.095|22 & 0.23|26 & 0.18|30 & 0.36|23 & 0.44|14 \\
 \text{DAX vs. random} & 0.15|12 & -0.067|18 & -0.12|24 & 0.029|31 & 0.31|32 & 0.27|27 & 0.59|15\\
 \hline
\end{array}
$
\begin{center} Table 4. Comparison between the daily stock market sequences vs. an hypothesised log-normal accumulation of price directions.
\end{center}

\subsubsection{Correlation matrix: market vs. Turing machines}

\begin{center}
$
\begin{array}{|c|c|c|c|c|c|c|}
\hline
\text{market vs. TM}  & 5 & 6 & 7 & 8 & 9 & 10 \\
\hline
 \text{DJIA vs. TM} & 0.42|16 & 0.20|21 & 0.42|24 & -0.021|35 & -0.072|36 & 0.20|47 \\
 \text{SP500 vs. TM} & 0.48|18 & 0.30|24 & -0.070|32 & 0.32|39 & 0.26|47 & 0.40|55 \\
 \text{NASDAQ vs. TM} & 0.67|17 & 0.058|25 & 0.021|32 & 0.26|42 & 0.076|49 & 0.17|57 \\
 \text{FTSE350 vs. TM} & 0.30|17 & 0.39|22 & 0.14|29 & 0.43|36 & 0.013|41 & 0.038|55 \\
 \text{CAC40 vs. TM}  & 0.49|17 & 0.026|25 & 0.41|32 & 0.0056|38 & 0.22|47 & 0.082|56 \\
\hline
\end{array}
$
\end{center}
\begin{center}
Table 5. The comparison to TM revealed day lengths better correlated than other, although\\ their significance remained weak and unstable, with a tendency, however, to positive correlations.
\end{center}

\subsubsection{Correlation matrix: market vs. cellular automata}

\begin{flushleft}
$
\begin{array}{|c|c|c|c|c|c|c|c|}
\hline
 \text{market vs. CA} & 4 & 5 & 6 & 7 & 8 & 9 & 10 \\
\hline
 \text{DJIA vs. CA} & -0.14|16 & 0.28|32 & -0.084|63 & -0.049|116 & 0.10|148 & 0.35|111 & 0.51|59 \\
 \text{SP500 vs. CA} & -0.16|16 & 0.094|32 & 0.0081|64 & 0.11|116 & 0.088|140 & 0.17|117 & 0.40|64 \\
 \text{NASDAQ vs. CA} & 0.065|16 & 0.25|32 & 0.19|63 & 0.098|116 & 0.095|148 & 0.065|131 & 0.36|65 \\
 \text{FTSE350 vs. CA} & -0.16|16 & -0.15|32 & 0.12|64 & -0.013|120 & -0.0028|146 & 0.049|124 & 0.42|76 \\
 \text{CAC40 vs. CA} & -0.035|16 & 0.36|32 & 0.21|64 & 0.064|114 & 0.20|138 & 0.25|114 & 0.33|70\\
\hline
\end{array}
$
\end{flushleft}
\begin{center} Table 6. When compared to the distribution from cellular automata, the correlation was greater. Each column had pairs of score means: (-0.09, 16), (0.17, 32), (0.09, 64), (0.042, 116), (0.096, 144), (0.18, 119), (0.41, 67) for 4 to 10 days, for which the last 2 (9 and 10 days long) have significant levels of correlation according to their critical values and the number of elements compared.
\end{center}

\subsection{Backtesting}
\label{backtesting}

Applying the same methodology over a period of a decade, from 1980 to 1990 (old market), to three major stock markets for which we had data for the said period of time, similar correlations were found across the board, from weak to moderately weak--- though the trend was always toward positive correlations.

\subsubsection{Correlation matrix: old market vs. CA distribution}
$
\begin{array}{|c|c|c|c|c|c|c|c|}
 \hline
 \text{old market vs. CA}  & 4 & 5 & 6 & 7 & 8 & 9 & 10 \\
 \hline
 \text{DJIA vs. CA} & 0.33|10 & 0.068|16 & 0.51|21 & 0.15|28 & -0.13|31 & 0.12|32 & 0.25|29 \\
 \text{SP500 vs. CA} & 0.044|13 & 0.35|19 & 0.028|24 & 0.33|33 & 0.45|33 & 0.00022|30 & 0.37|34 \\
 \text{NASDAQ vs. CA} & 0.45|10 & 0.20|17 & 0.27|24 & 0.16|30 & 0.057|31 & 0.11|34 & 0.087|32\\
 \hline
\end{array}
$
\begin{center}
\noindent Table 7. Comparison matrix of frequency distributions of daily price directions of three stock markets from 1980 to 1990.
\end{center}

The distributions indicate that price changes are unlikely to rise by more than a few points for more than a few days, while greater losses usually occur together and over longer periods. The most common sequences of changes are alternations. It is worth noticing  that sequences are grouped together by reversion and complementation relative to their frequency, whereas traditional probability would have them occur in no particular order and  with roughly the same frequency values.

Tables 8 and 9 illustrate the kind of frequency distributions from the stock markets (in this case for the DJI) over tuples of length 3 with which distributions from the market data were compared with and its statistical correlation evaluated section \label{experiments} between four other stock markets and over larger periods of time up to 10 closing daily prices.

\begin{center}
$\begin{array}{|c|r|}
\hline
tuple&prob.\\
\hline
000&0.139\\
001&0.130\\
111&0.129\\
011&0.129\\
100&0.129\\
110&0.123\\
101&0.110\\
010&0.110\\
\hline
\end{array}
$
\medskip \\Table 8. 3-tuples distribution from the DJI price difference time series for the past 80 years. 1 means that a price rose, 0 that it fell or remained the same as described in the construction of the binary sequence $b_n$ as described in \ref{detecting} and partitioned in 3-tuples for this example. By rounding to the nearest multiple of .4 (i.e. dismissing decimal fraction price variations of this order) some more stable patterns start to emerge.
\end{center}

\text{\\}

\begin{center}
$\begin{array}{|c|c|}
\hline
tuple&prob.\\
\hline
000& 0.00508\\
111& 0.00508\\
001& 0.00488\\
011& 0.00488\\
100 & 0.00488\\
110 & 0.00468\\
010 & 0.00483\\
101 & 0.00463\\
\hline
\end{array}
$
\medskip \\Table 9. 3-tuples from the output distribution produced by running all 4-state\\ 2-symbol Turing machines starting from an empty tape first on a background of 0s and then running it again on a background of 1s to avoid asymmetries due to the machine formalism convention.
\end{center}

The algorithmic model predicts a greater incidence of simple signatures as trends under the \emph{noise} modelled by the Brownian motion model, such as signatures 000\ldots of price stability. It also predicts that random-looking signatures of higher volatility will occur more if they are already occurring, a signature in unstable times where the Brownian motion no longer works in these kind of events outside the main Bell curve. Our empirical samples show that given the weak to strong correlations, it is indeed the case that a small component of the price phenomenon in financial markets may follow rules, and that the upshot may be the hidden rules and trends underlying and driving  the market.

\section{Further considerations}
\label{furtherconsiderations}

\subsection{Rule-based agents}

For sound reasons, economists are used to standardising their discussions by starting out from certain basic assumptions. One common assumption in economics is that actors in the market are decision makers, often referred to as \emph{rational agents}. According to this view, rational agents make choices by assessing  possible outcomes and assigning a utility to each in order to make a decision. In this rational choice model, all decisions are arrived at by a \emph{rational} process of weighing costs against benefits, and not randomly.

An agent in economics or a player in game theory is an actor capable of decision making. The idea is that the agent initiates actions, given the available information, and tries to maximise his or her chances of success (traditionally their personal or collective utilities), whatever the ultimate goal may be. The algorithm that each agent takes may be non-deterministic, which means that the agent may make decisions based on probabilities, not that at any stage of the process a necessarily truly random choice is made. It actually doesn't matter much whether their actions may be perceived as mistaken, or their utility questioned. What is important is that agents follow rules, or if any chance is involved there is another large part in it not random at all (specially when one takes into consideration the way algorithmic trading is done). The operative assumption is that the individual has the cognitive ability to weigh every choice he/she makes, as opposed to taking decisions stochastically. This is particularly true when there is nothing else but computers making the decisions.

This view, wherein each actor can be viewed as a kind of automaton following his or her own particular rules, does not run counter to the stance we adopt here, and it is in perfect agreement with the algorithmic approach presented herein (and one can expect the market to get more algorithmic as more automatization is involved). On the contrary, what we claim is that if this assumption is made, then the machinery of the theory of computation can be applied, particularly the theory of algorithmic information (AIT). Hence market data can be said to fall within the scope of algorithmic probability.

Our approach is also compatible with the emergent field of behavioural economics\cite{camerer}, provided the set of cognitive biases remain grounded in rules. Rules followed by emotional (or non-rational) traders can be as simple as imitating behaviour, repeating from past experience, acting out of fear, taking advice from others or following certain strategy. All these are algorithmic in nature in that they are rule based, despite their apparent idiosyncrasy (assuming there are no real clairvoyants with true metaphysical powers). Even though they may look random, what we claim, on the basis of algorithmic probability and Levin's distribution, is that most of these behaviours follow simple rules. It is the accumulation of simple rules rather than the exceptional complicated ones  which actually generate trends.

If the market turns out to be based on simple rules and driven by its intrinsic complexity rather than by the action of truly random external events, the choice or application of rational theory would be quite irrelevant. In either case, our approach remains consistent and relevant. Both the rational and, to a large extent, the behavioural agent assumptions imply that what we are proposing here is that algorithmic complexity can be directly applied to the field of market behaviour, and that our model comes armed with a natural toolkit for analysing the market, viz. algorithmic probability.

\subsection{The problem of over-fitting}

When looking at a set of data following a distribution, one can claim, in statistical terms, that the source generating the data is of the nature that the distribution suggests. Such is the case when a set of data follows a model, where depending on certain variables, one can say with some degree of certitude that the process generating the data follows the model.

It seems to be well-known and largely accepted among economists that one can basically fit anything to anything else, and that this has shaped most of the research in the field, producing a sophisticated toolkit dictating how to achieve this fit as well as how much of a fit is necessary for particular purposes, even though such a fit may have no relevance either to the data or to particular forecasting needs, being merely designed to produce an instrument with limited scope fulfilling a specific purpose. 

However, a common problem is the problem of over-fitting, that is, a false model that may fit perfectly with an observed phenomenon. A statistical comparison cannot actually be used to categorically prove or disprove a difference or similarity, only to favour one hypothesis over another.

To mention one of the arbitrary parameters that we might have taken, there is the chosen rounding. We found it interesting that the distributions from the stock markets were sometimes unstable to the rounding process of prices. Rounding to the closest .4 was the threshold found to allow the distribution to stabilise. This instability may suggest that there are two different kinds of forces acting, one producing very small and likely negligible price movements (in agreement to the random walk expectation), and other producing the kind of qualitative changes in the direction of prices that we were interested in. In any case this simply results in the method only being able to predict changes of the order of magnitude of the rounding proceeding from the opposite direction, assuming that the data is not random, unlike the stochastic models.

As proven by Levin and Solomonoff, the algorithmic probability measure (the universal distribution) will outperform any other, unless other information is available that helps to foresee the outcome, in which case an additional variable could be added to the model to account for this information. But since we've been suggesting that information will propagate fast enough even though the market is not stochastic in nature, deleting the patterns and making them unpredictable, any additional assumption only complicates the model. In other words, Levin's universal distribution is optimal over all non-random distributions\cite{levin2}, in the sense that the algorithmic model is by itself the simplest model fitting the data when this data is produced by a process (as opposed to being randomly generated). The model is itself ill-suited to an excess of parameters argument because it basically assumes only that the market is governed by rules.

Algorithmic probability rests upon two main principles: the principle of multiple explanations, which states that one should keep all hypotheses that are consistent with the data, and a second principle known as Occam's razor, which states that when inferring causes, entities should not be multiplied beyond necessity, or, alternatively, that among all hypotheses consistent with the observations, the simplest should be favoured. As for the choice of an a priori distribution over a hypothesis, this amounts to assigning simpler hypotheses a higher probability and more complex ones a lower probability. So this is where the concept of algorithmic complexity comes into play.

As proven by Solomonoff and Levin, any other model will simply overlook some of the terms of the algorithmic probability sum. So rather than being more precise, any other model will differ from algorithmic probability in that it will necessarily end up overlooking part of the data. In other words, there is no better model taking into account the data than algorithmic probability. As Solomonoff has claimed, one can't do any better. Algorithmic inference is a time-limited optimisation problem, and algorithmic probability accounts for it simply.

\newpage

\section{Conclusions and further work}
\label{conclusions}

When looking at a large-enough set of data following a distribution, one can in statistical terms safely assume that the source generating the data is of the nature that the distribution suggests. Such is the case when a set of data follows a normal distribution, where depending on certain statistical variables, one can,  for example, say with a high degree of certitude that the process generating the data is of a random nature. If there is an algorithmic component in the empirical data of price movements in financial markets, as  might be suggested by the distribution of price movements, algorithmic information theory may account for the deviation from log-normality as argued herein. In the words of Velupillai\cite{vela}---quoting Clower\cite{clower} talking about Putnam's approach to a theory of induction\cite{putnam2}\cite{putnam}---\emph{This may help ground `economics as an inductive science'} again. 

One may well ask whether a theory which assumes that price movements follow an algorithmic trend ought not to be tested in the field to see whether it outperforms the current model. The truth is that the algorithmic hypothesis would easily outperform the current model, because it would account for recurrent periods of instability. The current theory, with its emphasis on short term profits, is inclined to overlook these, for reasons that are probably outside the scope of scientific inquiry. In our understanding, the profits attributed to the standard current model are not really owed to the  model as such, but rather to the mechanisms devised to control the risk-taking inspired by  the overconfidence that the model generates.

In a strict sense, this paper describes the ultimate possible numerical simulation of the market when no further information about it is known (or cannot be known in practise) assuming no other (neither efficient markets nor general equilibrium) but actors following a set of rules and therefore to behave algorithmically at least at some extent hence potentially modelled by algorithmic probability. 

The simulation may turn out to be of limited predictive value---for looking no more than a few days ahead and modelling weak signals---due to the deleting patterns phenomenon (i.e. the time during which the market assimilates new information). More experiments remain to be done which carefully encode and take into consideration other variables, such as the magnitude of prices, for example, looking at consecutive runs of gains or loses.

\section*{Acknowledgments}

Hector Zenil wishes to thank Vela Velupillai and Stefano Zambelli for their kind invitation to take part in the workshop on Nonlinearity, Complexity and Randomness at the Economics Department of the University of Trento, and for their useful comments. Jason Cawley, Fred Meinberg, Bernard Fran\c{c}ois and Raymond Aschheim provided helpful suggestions, for which many thanks. Any misconceptions remain of course the sole responsibility of the authors.

\end{document}